\documentclass[a4paper,showpacs,prb,twocolumn]{revtex4}
\usepackage{graphicx}
\usepackage{dcolumn}

\begin{document}

\title{Interacting electrons in the Aharonov-Bohm interferometer}
\author{S. Ihnatsenka}
\author{I. V. Zozoulenko}
\affiliation{Solid State Electronics, Department of Science and Technology, Link\"{o}ping
University, 60174 Norrk\"{o}ping, Sweden}
\date{\today}

\begin{abstract}
We present a microscopic picture of quantum transport in the Aharonov-Bohm (AB)
interferometer taking into account electron interaction within the Hartree and the spin
density functional theory approximations. We discuss the structure of the edge states for
different number of the Landau levels in the leads, their coupling to the states in the
central island and the formation of compressible/incompressible strips in the
interferometer. Based on our results we discuss the existing theories of the unexpected
AB periodicity, which essentially rely on specific phenomenological models of the states
and their coupling in the interferometer. Our work provides a basis for such the
theories, giving a detailed microscopic description of the propagating states and the
global electrostatics in the system at hand.
\end{abstract}

\maketitle

\section{Introduction}

The recent years have witnessed a renewed interest in studies of
magnetotransport in quantum Hall systems in confined geometries\cite%
{Goldman_frac,Goldman_2005,Goldman_2007,Rosenow,DasSarmaPRL05,5/2_nature_physics,top_com_RMP}%
. These studies are motivated in part by the prospect of topological quantum
computing\cite{DasSarmaPRL05,5/2_nature_physics,top_com_RMP} as well as by
fundamental interest to explore novel exciting physics related to e.g.
exotic fractional statistics in two dimensional systems\cite{Goldman_frac}.
Some recent studies have revealed new unexpected features in systems that
have been extensively studied in the nighties and that seemed to be well
understood since long time ago. This includes, for example, an unexpected
periodicity of the Aharonov-Bohm (AB) electron interferometer in the integer
quantum Hall regime of the edge state transport revealed in the experiments
of Camino \textit{et al}.\cite{Goldman_2005,Goldman_2007}

\begin{figure}[tbp]
\includegraphics[scale=1.0]{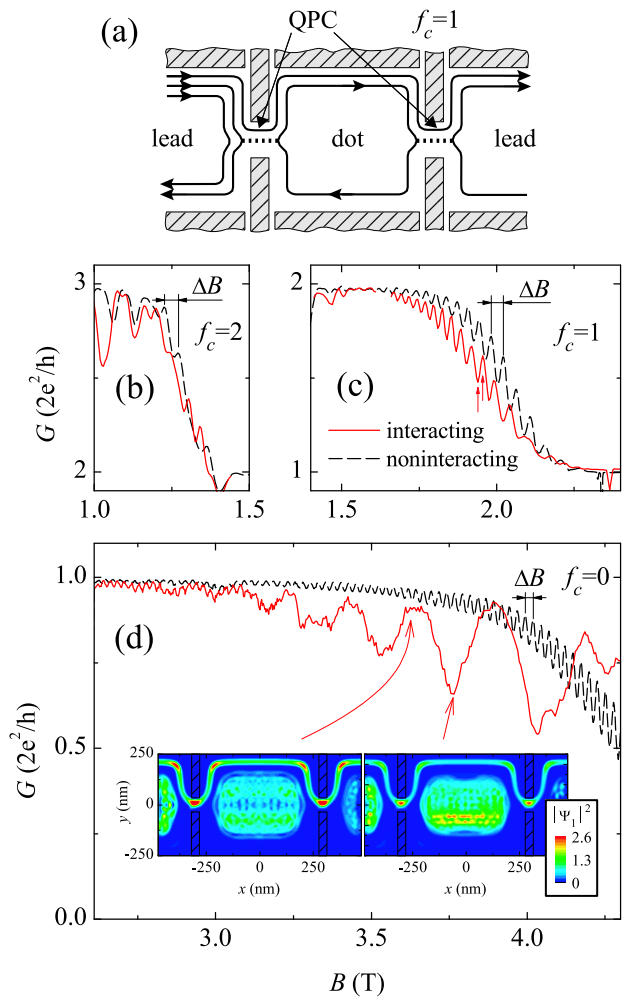}
\caption{(color online) (a) Schematic geometry of the AB interferometer.
Shaded regions corresponds to the metallic gates on the top of the GaAs
heterostructure. The geometrical size of the dot is $500\times520$ nm$^2$.
The diagram illustrates the case of $f_c=1$ and $f_{lead}=3$ corresponding
to one fully transmitting states through the constrictions and 3 propagating
states in the leads (for the calculated wave function for this case see Fig.
\protect\ref{fig:psi}). (b), (c), (d) The AB oscillations for respectively $%
f_c=2, 1, 0$ calculated for interacting and noninteracting electrons (solid
and dashed lines respectively). $\Delta B$ shows the expected periodicity
according to Eq. (\protect\ref{period_1}). The arrows in (c) indicate the
magnetic fields corresponding to calculated LDOS shown in Fig. \protect\ref%
{fig:LDOS}. The inset in (d) illustrates the wave function due to the first
propagating state in the leads.}
\label{fig:AB_oscillations}
\end{figure}

A typical Aharonov-Bohm interferometer\cite%
{Wees,Alphenaar,Taylor,Field,Goldman_2005,Goldman_2007} includes an electron
island coupled to the leads by two quantum point contacts (QPC\textbf{s}),
see Fig. \ref{fig:AB_oscillations} (a). At a given magnetic field there are $%
f_{leads}$ propagating edge states in the leads. The electron density in the
constrictions of the QPC is smaller than in the leads, and hence only the
lowest $f_{c}$ states are fully transmitted through the constriction,
whereas the remaining highest $f_{leads}-f_{c}$ states are partially or
fully reflected. A typical conductance of the AB interferometer as a
function of magnetic field exhibits a step-like structure with plateaus
separated by wide transitions regions\cite%
{Wees,Alphenaar,Taylor,Field,Goldman_2005,Goldman_2007}. This structure of
the conductance reflects successive depopulation of the magnetosubbans in
the constrictions. The plateau regions correspond to the field regions where
the QPC openings are fully transparent (the transmission coefficient through
an individual QPC is integer, $T\cong f_{c}$), and transition regions
between these plateaus correspond to the partial transparent QPC openings
(the transmission coefficient is non-integer, $f_{c}<T<f_{c}+1)$. In the
latter case, reflection on the QPC openings inside the island confines the
partially transmitted ($f_{c}+1$)-th state between two QPCs which gives rise
to pronounced AB conductance oscillations in the transition regions between
the plateaus. Note that in the weak coupling regime when the number of
propagating states in the constrictions is reduced below one, $f_{c}=0,$ the
AB oscillations are suppressed by the single electron charging effects.\cite%
{Alphenaar,Taylor,Field} In this case the charge of the electron island
inside interferometer becomes quantized, and the conductance exhibits
familiar Coulomb blockade (CB) peaks corresponding to addition/removing one
electron to/from the interior of the central island.

According to the conventional theory of the Aharonov-Bohm interferometer its
conductance shows a peak each time the enclosed flux $\phi =BS$ changes by
the flux quantum $\phi _{0}=h/e$, $\Delta (BS)=\phi _{0}$.\cite{Davies_book}
Thus, the conductance of the interferometer as a function of the magnetic
field exhibits the periodicity
\begin{equation}
\Delta B=\frac{\phi _{0}}{S},  \label{period_1}
\end{equation}%
with $S$ being the area of the island. The inclosed flux through the
interferometer can also be varied at a fixed magnetic field by changing a
gate voltage. In the case when the area changes linearly with the change in
the gate voltage, $\Delta S=\alpha \Delta V_{g},$ the expected periodicity
is
\begin{equation}
\Delta V_{g}=\phi _{0}/\alpha B.  \label{period_Vg}
\end{equation}

The first experimental study of the AB interferometer in lateral GaAs
heterostructures was performed by van Wees \textit{at al}.\cite{Wees} They
reported a good agreement between the theory and experiment with some
deviation from Eq. (\ref{period_1}) for the case of several propagating
modes in the constrictions ($f_{c}\geq 2$). They attributed this deviation
to the effect of magnetic field on the location of the edge states
corresponding to different $f_{c}$. Since these pioneering experiments, the
interpretation of the AB oscillations based on Eq. (\ref{period_1}) has been
widely accepted. However, in recent experiments of Camino \textit{et al.}%
\cite{Goldman_2005,Goldman_2007} the validity of the conventional theory of the AB
oscillations in lateral semiconductor heterostructures has been questioned. In
particularly, Camino \textit{et al.} demonstrated that periodicity of the AB oscillations
as a function of the magnetic field depends of the number of fully transmitted states in
the constriction $f_{c}$ and is well described by the dependence
\begin{equation}
\Delta B=\frac{1}{f_{c}}\frac{\phi _{0}}{S},  \label{period_f_c}
\end{equation}%
which obviously differs by a factor $1/f_{c}$ from the conventional formula (%
\ref{period_1}). On the other hand, the back-gate charge period is the same
(one electron) for all $f_c$, independent of on the magnetic field in stark
contrast to Eq. (\ref{period_Vg}). Moreover, Camino \textit{et al. }\cite%
{Goldman_2005} re-analyzed the data of the experiment of van Wees \textit{at
al}.\cite{Wees} and concluded that it is, within the experimental uncertainty, also
described by Eq. (\ref{period_f_c}). (Note that the same re-interpretation of the van
Wees \textit{et al}. experiment was first proposed by Dharma-wardana \textit{et
al}.\cite{Dharma-wardana})

It should be stressed that Eqs. (\ref{period_Vg})-(\ref{period_f_c}) represent a
significant departure from the conventional description of the AB oscillations based on
Eq. (\ref{period_1}). The latter relies on a one-electron picture of non-interacting
electrons, whereas the former require accounting for electron interaction and/or CB
charging effects. An interplay between the AB and CB oscillations has been experimentally
studied by Taylor \textit{et al}.\cite{Taylor} and Field \textit{et al}.\cite{Field}
Taylor \textit{et al}.\cite{Taylor} has established a simple condition for the onset of
the CB oscillations in their structure, namely the dot has to be in the weak coupling
regime with only partially transmitted states in the constrictions, $f_{c}=0.$ On the
contrary, Field \textit{et al.}\cite{Field} found coexistence of the AB and CB
oscillations extended even into the open dot regime when $f_{c}\geq 2$. A persistence of
the Coulomb blockade oscillations into the open regime $f_{c}\geq 1$ was also discussed
by Alphenaar \textit{et al.}\cite{Alphenaar}. Note that possibility of the Coulomb
charging effects in the strongly coupled regime has been a subject of interesting
discussions for the case of quantum antidots in the integer quantum Hall
regime\cite{adot} as well as open quantum dots at zero magnetic field\cite{Liang}. It
should be also mentioned that similar deviations from the standard AB formula that are
also well described by Eq. (\ref{period_f_c}) have been very recently reported by Goldman
\textit{et al.}\cite{Goldman_2008} for the case of an \textit{antidot-based} AB
interferometer.

The effect of electron interaction and Coulomb charging on the conductance of the AB
interferometer in the open regime of $f_{c}\geq 1$ was studied by Dharma-wardana
\textit{et al.}\cite{Dharma-wardana} and very recently by Rosenow and
Halperin\cite{Rosenow}. Using different approaches they both arrived to the same
conclusion that the AB oscillations can be modulated by CB-type effect leading to the
novel periodicity of the oscillations described by Eq. (\ref{period_f_c}). However, their
models have been based on very different microscopic mechanisms of interaction and
charging in the
interferometer. In the model of Dharma-wardana \textit{et al.}\cite%
{Dharma-wardana} the predicted modulation is due to the enhanced screening
of the usual CB oscillations by fully transmitted states through
constriction effectively acting as metallic strips. In contrast, the
predictions of Rosenow and Halperis\cite{Rosenow} are based on the
assumption of the coupling between the states in the leads and the central
compressible island inside the interferometer.

Thus, understanding of the role of electron interaction and charging in the
AB interferometer and identification of the origin of the unexpected
periodicity of the oscillations (\ref{period_f_c}) require detailed
knowledge of the microscopic structure as well as the coupling strength
between different states in the leads and in the central island. To the best
of our knowledge such calculations have not been reported in the literature
yet. At the same time, this information is essential in theories like those
developed in Refs. [\onlinecite{Dharma-wardana,Rosenow}] that rely on
specific models of the coupling between states in the interferometer.

In the present paper we perform such the calculations for the AB interferometer in the
integer quantum Hall regime where electron interaction and spin effects are included
within the spin-density functional theory (DFT). The utilized approach corresponds to the
first-principles magnetoconductance calculations (within the effective mass
approximation) that start from a geometrical layout of the device, are free from
phenomenological parameters of the theory, and do not rely on model
Hamiltonians.\cite{open_dot_PRB,QPC} The power of this approach has been recently
demonstrated for related systems (quantum dots and quantum wires) where a
\textit{quantitative} agreement with the corresponding experiments has been
achieved\cite{narrow_wires,PRL}.

The paper is organized as follows. A brief description of the model is given
in Sec II. Section III presents results for the edge state structure and
coupling between states in the leads and in the dot calculated within the
Hartree approximation as well as within the spin-DFT approach. We discuss
the obtained results and outline their relation to the experiment and
exising theories in Sec. IV. The conclusion is presented in Sec. V.

\section{Model.}

\label{sec:Model}

We consider an electron interferometer defined by split-gates in the GaAs heterostructure
in an open quantum dot geometry similar to those studied
experimentally\cite{Goldman_2005,Goldman_2007,Wees,Alphenaar,Taylor,Field},
see Fig. \ref{fig:AB_oscillations}. The geometrical size of the dot is $%
500\times 520$ nm$^{2}$, the geometrical width of the QPC openings is 80 nm,
and the distance from the two-dimensional electron gas to the surface is $%
b=50$ nm. The Hamiltonian of the whole system (the island + the
semi-infinite leads) in the framework of the density-functional theory (DFT)
within the Kohn-Sham formalism\cite{Giuliani_Vignale} can be written in the
form $H=H_{0}+V(\mathbf{r}),$ where
\begin{equation}
H_{0}=-\frac{\hbar ^{2}}{2m^{\ast }}\left\{ \left( \frac{\partial }{\partial
x}-\frac{eiBy}{\hbar }\right) ^{2}+\frac{\partial ^{2}}{\partial y^{2}}%
\right\}
\end{equation}%
is the kinetic energy in the Landau gauge, and the total confining potential
\begin{equation}
V(\mathbf{r})=V_{conf}(\mathbf{r})+V_{H}(\mathbf{r})+V_{xc}^{\sigma }(%
\mathbf{r})+V_{Z},
\end{equation}%
where $V_{conf}(\mathbf{r})$ is the electrostatic confinement (including
contributions from the top gates, the donor layer and the Schottky barrier),
$V_{H}(\mathbf{r})$ is the Hartree potential,
\begin{equation}
V_{H}(\mathbf{r})=\frac{e^{2}}{4\pi \varepsilon _{0}\varepsilon _{r}}\int d%
\mathbf{r}\,^{\prime }n(\mathbf{r}^{\prime })\left( \frac{1}{|\mathbf{r}-%
\mathbf{r}^{\prime }|}-\frac{1}{\sqrt{|\mathbf{r}-\mathbf{r}^{\prime
}|^{2}+4b^{2}}}\right) ,  \label{V_H}
\end{equation}%
where $n(\mathbf{r})$ is the electron density, the second term corresponds
to the mirror charges situated at the distance $b$ from the surface, $%
\varepsilon _{r}=12.9$ is the dielectric constant of GaAs. $V_{xc}^{\sigma }(%
\mathbf{r})$ is the exchange-correlation potential in the local spin-density
approximation where $\sigma $ stands for spin-up, $\uparrow $, and
spin-down, $\downarrow $, electrons, and $V_{Z}$ is a standard Zeeman term.
In calculation of $V_{xc}^{\sigma }(\mathbf{r})$ we utilized a commonly used
parametrization of Tanatar and Ceperly\cite{TC}. (A detailed description of
the Hamiltonian can be found in Refs. \onlinecite{open_dot_PRB,QPC}). The
dot and the leads are treated on the same footing, i.e. the electron
interaction and the magnetic field are included both in the lead and in the
dot regions. In what follows we will mostly concentrate on the Hartee
approximation (i.e. when $V_{xc}^{\sigma }(\mathbf{r})=0).$ This is because
the main conclusions concerning the Aharonov-Bohm oscillations in the system
at hand are qualitatively similar for the spinless Hartree case ($%
V_{xc}^{\sigma }(\mathbf{r})=0$) and the spin-resolved DFT case ($%
V_{xc}^{\sigma }(\mathbf{r})\neq 0$).

We calculate the self-consistent electron densities, potentials and the conductance on
the basis of the Green's function technique. The description of the method can be found
in Refs. \onlinecite{open_dot_PRB,QPC} and thus the main steps in the calculations are
only briefly sketched here. First we compute the self-consistent solution for the
electron density, effective potential and the Bloch states in the semi-infinite leads by
the technique described in Ref. \onlinecite{wires}. Knowledge of the Bloch states allows
us to find the surface Greens function of the semi-infinite leads. We then calculate the
Green's function of the central section of the structure by adding slice by slice and
making use of the Dyson equation on each iteration step. Finally we apply the Dyson
equation in order to couple the left and right leads with the central section and thus
compute the full Green's function $\mathcal{G}^{\sigma }(E)$ of the whole system. The
electron density is integrated from the Green's function (in the real space),
\begin{equation}
n^{\sigma }=-\frac{1}{\pi }\int_{-\infty }^{\infty }\Im \lbrack \mathcal{G}%
^{\sigma }(\mathbf{r},\mathbf{r},E)]f_{FD}(E-E_{F})dE,
\end{equation}%
where $f_{FD}$ is the Fermi-Dirac distribution. This procedure is repeated
many times until the self-consisten solution is reached; we use a
convergence criterium $\left\vert n_{i}^{out}-n_{i}^{in}\right\vert
/(n_{i}^{out}+n_{i}^{in})<10^{-5}$, where $n_{i}^{in}$ and $n_{i}^{out}$ are
input and output densities on each iteration step $i.$

Finally the conductance is computed from the Landauer formula, which in the
linear response regime is
\begin{equation}
G^{\sigma }=-\frac{e^{2}}{h}\int_{-\infty }^{\infty }dET^{\sigma }(E)\frac{%
\partial f_{FD}(E-E_{F})}{\partial E},
\end{equation}%
where the transmission coefficient for the spin channel $\sigma ,$ $T^{\sigma
}(E)$ , is calculated from the Green's function between the leads. \cite%
{open_dot_PRB,QPC} All the calculations reported in the present paper are
performed for the temperature $T=0.2$K.

To outline the role of the electron interaction we also calculate the
conductance of the open dot in the Thomas-Fermi (TF) approximation where the
self-consistent electron density and potential are given by the standard TF
equation at zero field. This approximation does not capture effects related
to electron-electron interaction in quantizing magnetic field such as
formation of compressible and incompressible strips and hence it corresponds
to noninteracting one-electron approach where, however, the total
confinement is given by a smooth fixed realistic potential, see Refs. %
\onlinecite{open_dot_PRB,QPC} for details.

In order to provide correct interpretation of the results reported in this paper, it is
important to outline the validity and limitations of the present approach. Our
calculations correspond to a so-called \textquotedblleft standard
approach\textquotedblright \cite{Koentopp} based on the ground-state DFT in the Landauer
formula. It has been demonstrated that this approach accurately describes the conductance
in the regime of the strong coupling when the conductance of the QPCs connecting the
device region and the leads exceeds the conductance unit $G_{0}$($=2e^{2}/h$ for the
spin-degenerate electrons). This corresponds to the case when charge quantization inside
the device is not expected to occur. In this regime the \textquotedblleft standard
approach\textquotedblright\ was shown to reproduce not only qualitatively, but in many
cases even quantitatively the observed conductance of metallic nanowires\cite{Koentopp}
as well as GaAs lateral heterostructures\cite{narrow_wires,PRL}.

However, the reliability of this approach has been questioned for the case
of the weak coupling where the QPC conductance drops below the conductance
unit $G_{0}$ such that charge inside the device becomes quantized (i.e. in
the Coulomb blockade regime)\cite{Toher,Koentopp,QPC}. This is due to the
uncorrected self-interaction errors in the standard DFT approach (related to
the lack of the derivative discontinuity in the exchange-correlation
potential) for the case when localization of charge is expected to occur.
Because of this, we do not expect the present approach to provide a reliable
conductance for the case of the weak coupling $f_{c}=0$ (Fig. \ref%
{fig:AB_oscillations} (d)), where the experiments exhibit the Coulomb
blockaded conductance\cite{Alphenaar,Taylor}.

While the present approach is not expected to account for single-electron tunneling in
the conductance (leading to the Coulomb blockade peaks), one can expect that it correctly
reproduces a global electrostatics of the interferometer and microscopic structure of the
quantum mechanical edge states regardless whether the conductance is dominated by a
single-electron charging or not. This is because the interferometer contains a large
number of electrons, $\sim 400-500,$ and thus the electrostatic charging caused by a
single electron hardly affects the total confining potential of the interferometer. Thus
the results of the present study provide an accurate information concerning the locations
of the propagating states and the structure of compressible/incompressible strips in the
interferometer. Our calculations are also expected to provide a detailed information
concerning the coupling strengths between the states in the leads and in the island.

\section{Results}

\label{sec:Results}

Figure \ref{fig:AB_oscillations} (b)-(d) shows the conductance of the AB interferometer
as a function of magnetic field for spin-degenerate interacting (Hartree) and
noninteracting (Thomas-Fermi) electrons for different numbers of fully propagating
channels $f_{c}$ in the QPC openings, $f_{c}=2,1,0.$ In these figures the voltages on the
gates defining the QPCs are set such that the constrictions accommodate $f_{c}$ fully
transmitted (lowest) Landau levels, while the ($f_{c}+1)$-th Landau level is only
partially transmitted.

\begin{figure}[tbp]
\includegraphics[keepaspectratio,width=\columnwidth]{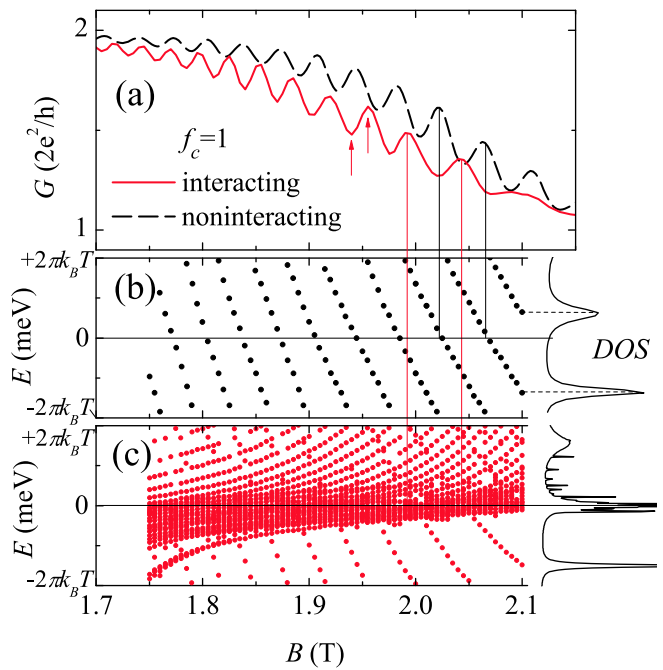} 
\caption{(color online) (a) The magnetoconductance of the AB interferometer
for $f_c=1$ for interacting and noninteracting spinless electrons (The
arrows in (a) indicate the magnetic fields corresponding to calculated LDOS
shown in Fig. \protect\ref{fig:LDOS}). Evolution of the resonant energy
levels in the vicinity of $E_F$ for (b) noninteracting and (c) interacting
electrons. The insets shows DOS in the dot for the specified value of the
field $B=2.1$ T; note that evolution of the energy levels was obtained from
the peak positions of the DOS at each given value of $B$.}
\label{fig:resonances}
\end{figure}

\begin{figure}[tbp]
\includegraphics[keepaspectratio, width=\columnwidth]{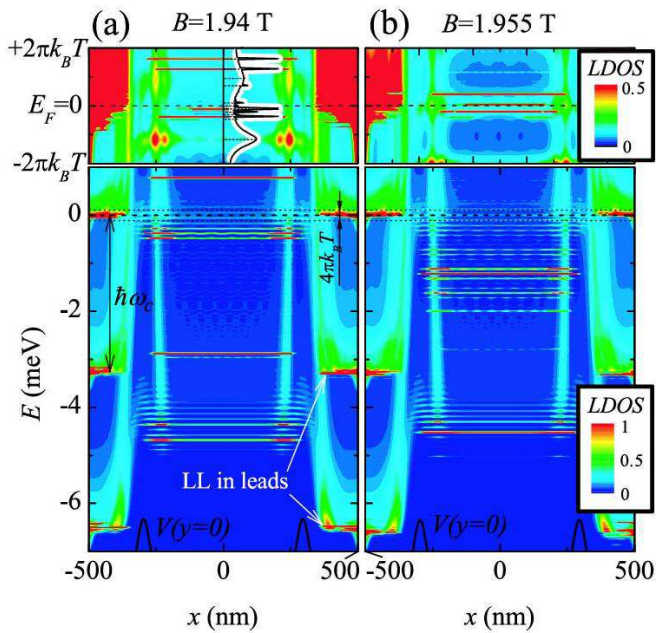}
\caption{(color online) The local density of states (LDOS) in the
interferometer for the case $f_{c}=1$ (spinless interacting electrons,
Hartree approximation). (a) and (b) correspond to a consecutive maximum and
minimum of the AB oscillations indicated by arrows in Fig. \protect\ref%
{fig:resonances}(a). The upper panels show resonant levels in the vicinity of
$E_{F}$ in an enlarged scale illustrating the resonant tunneling mediated by
the broad resonant levels of the quantum dot as well as showing the effect
of pinning of narrow resonances to $E_F$.}
\label{fig:LDOS}
\end{figure}

Let us first concentrate on the cases $f_{c}=1$ and $f_{c}=2$ when the
conductance shows the Aharonov-Bohm oscillations with the same periodicity
of $\Delta B=0.025$T. This periodicity is in excellent agreement with the
conventional AB formula (\ref{period_1}) where the actual dot area $S\approx
410\times 410$ nm$^{2}$ is slightly smaller than the geometric dot area $%
S_{act}=$ $500\times 520$ nm$^{2}$. The Aharonov--Bohm oscillations can be
related to evolution of the corresponding dot spectrum when a
single-electron states cross the Fermi level each time the flux through the
dot increases by the flux quantum. For the case of noninteracting electrons
this is illustrated in Fig. \ref{fig:resonances} (b) which shows an
evolution of the resonant levels as a function of magnetic field in the
vicinity of the Fermi energy. [To obtain the evolution of the resonant
levels we analyze the density of states (DOS) in the dot at each given $B$
and plot the positions of the peaks in the DOS as $B$ varies as illustrated
in Fig. \ref{fig:resonances} (b); see also illustration of the DOS and the
local density of states (LDOS) shown in Fig. \ref{fig:LDOS}]. Figure \ref%
{fig:resonances} (b) shows that the dot conductance exhibits a maximum each
time a resonant state sweeps past $E_{F}.$\ The resonant levels giving rise
to the AB oscillations are rather broad (with broadening $\Gamma \sim kT)$
because they are situated close to the dot boundaries and their coupling to
the states in the leads is rather strong.

\begin{figure}[tbp]
\includegraphics[keepaspectratio, width=\columnwidth]{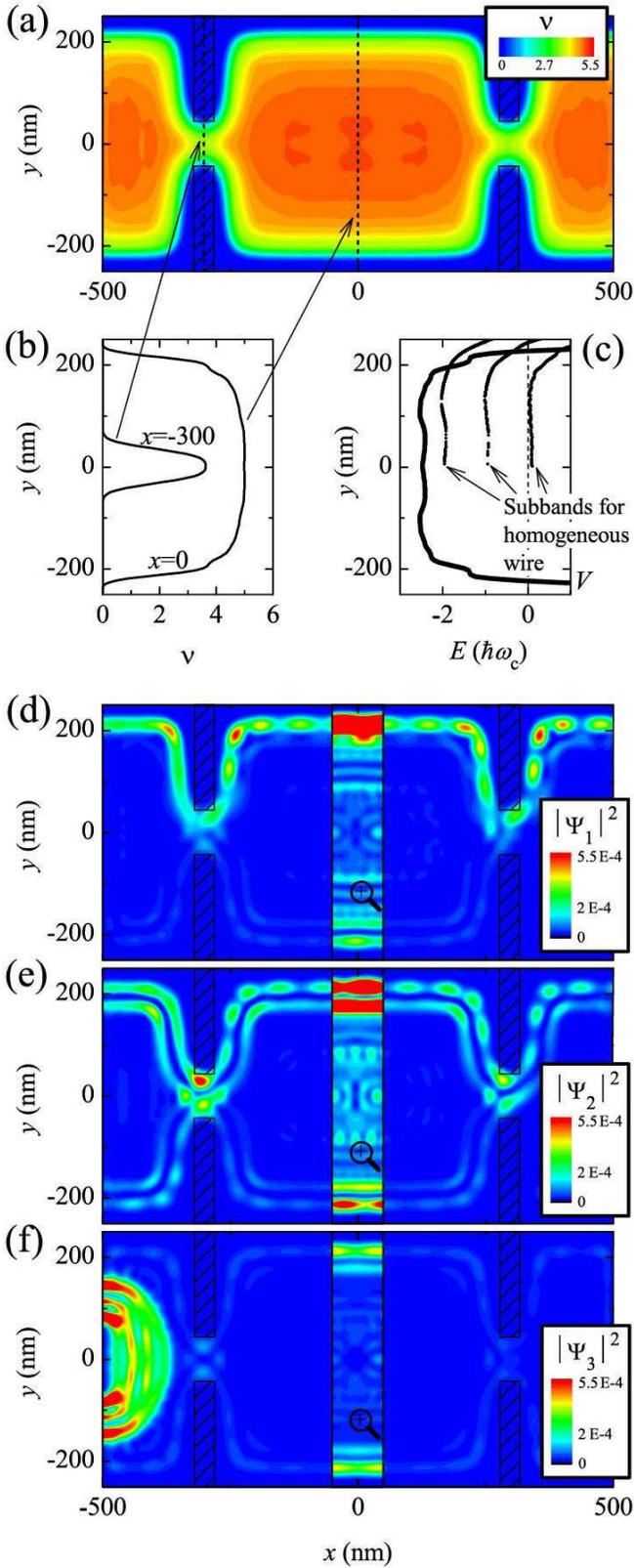} 
\caption{(color online) (a),(b) The local filling factor $\protect\nu (x,y)$
calculated for spinless interacting electrons (Hartree approximation) for $%
f_c=1$. (c) the total confining potential (thick line) and the corresponding
subband structure of an infinitive quantum wire. (d)-(f) the wave functions
modulus at $B=1.955$ T corresponding respectively to the first, second and
the third propagating states in the leads. The third subband being pinned at
the Fermi energy $E_{F}=0$ forms a compressible strip in the center of the
wire. The insets in (d)-(f) show strongly 10 times magnified intensity of
the wave functions in the middle section of the interferometer.}
\label{fig:psi}
\end{figure}

Figure \ref{fig:resonances} (c) shows an evolution of the resonant levels
for the case of interacting electrons. As for the case of noninteracting
electrons, the AB oscillations can be traced to relatively broad resonant
levels that sweep past the Fermi level each time the flux through the dot
increases by the flux quantum. However the resonant level structure of the
interacting electrons exhibits qualitatively new features. In addition to
the broad levels mediating the AB oscillations, the DOS shows many narrow
resonances concentrating near $E_{F}$. These resonances correspond to the
states residing inside the dot that are very weakly coupled to the leads
(hence small broadening). These states clearly show pinning to the Fermi
level (see Refs. \onlinecite{open_dot_PRB,PRL} for a detailed discussion of
the pinning effect in open quantum dots). However, the pinning of the inner
states to the Fermi energy does not imply that a compressible island forms
in the middle of the interferometer. Indeed, Fig. \ref{fig:LDOS} shows the
local density of states (LDOS) integrated in the transverse ($y$-direction).
In the magnetic field interval under consideration (corresponding to $%
f_{c}=1)$ there are three propagating states in the leads with the highest
one being always pinned to $E_{F}$ (see also Fig. \ref{fig:psi} (c)
depicting the magnetosubband structure in the leads). The dot itself shows
the Darwin-Fock type energy spectrum with a clear signature of the Landau
level (LL) condensation when the resonant levels concentrate around LLs of
the corresponding two-dimensional electron gas. Clearly, the upper bunch of
levels (concentrating around the second LL) is not pinned to the $E_{F}$.
The pinning of the highest LL to the $E_{F}$ (as well as the accompanying
formation of the compressible strip inside the dot) occurs at much higher
fields far above interval $f_{c}=1$. Thus the pinning of several resonant
levels to $E_{F}$ for the case $f_{c}=1$ shown in Figs. \ref{fig:resonances}
(c), \ref{fig:LDOS} represents an onset of formation of the compressible
island in the middle of the interferometer. It can be mentioned that a
related question whether the compressible strips form in an \textit{%
antidot-based} AB interferometer, has been a subject of recent debate\cite%
{adot_comp}.

\begin{figure}[tbp]
\includegraphics[scale=1.2]{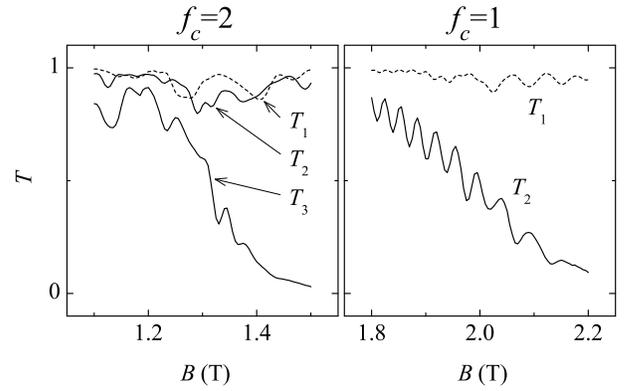}
\caption{Transmission coefficients $T_{i}=\sum_jT_{ji}$ from $i$-th mode in
the left lead to all available $j$-th modes in the right leads for
interacting spinless electrons (Hartree approximation) for (a) $f_c=2$ and
(b) $f_c=1$. (Note that corresponding total transmission is shown in Fig.%
\protect\ref{fig:AB_oscillations} (b), (c).)}
\label{fig:Tij}
\end{figure}

Despite of the difference in the structure of the DOS for interacting and
noninteracting electrons, their conductance is practically the same (the
small shift of the conductance curves relative to each other is due to a
small difference between the Hartree and TF densities). To understand the
reason for this, we inspect the wave functions in the interferometer, see
Fig. \ref{fig:psi}. We focus on the case of $f_{c}=1$ when the first state
in the leads $N_{lead}=1$ passes almost adiabatically through the QPC, the
third state $N_{lead}=3$ is reflected, and the AB oscillations are mostly
due to the second state $N_{lead}=2$ which is partially transmitted through
the QPC (see Fig. \ref{fig:Tij} illustrating the transmission coefficients
for $f_{c}=1$ and $f_{c}=2).$ In the field interval under consideration a
compressible strip in the leads forms only for the highest state $N_{lead}=3$
(a corresponding band structure for the lead is shown in Fig. \ref{fig:psi}%
). Two lowest states, $N_{lead}=1,2$, are not compressible and thus their
respective spatial location and structure are very similar for interacting
and noninteracting electrons. As a result, the coupling of these states to
the states in the dot are almost the same for interacting and noninteracting
electrons, and therefore the corresponding conductances are practically the
same. It should also be mentioned that in accordance to the discussion above
the wave function pattern does not show an evidence of the formation of the
compressible island in the center of the interferometer. In contrast, the
compressible strip corresponding to $N_{lead}=3$ is clearly seen in the
leads.

\begin{figure}[tbp]
\includegraphics[keepaspectratio, width=\columnwidth]{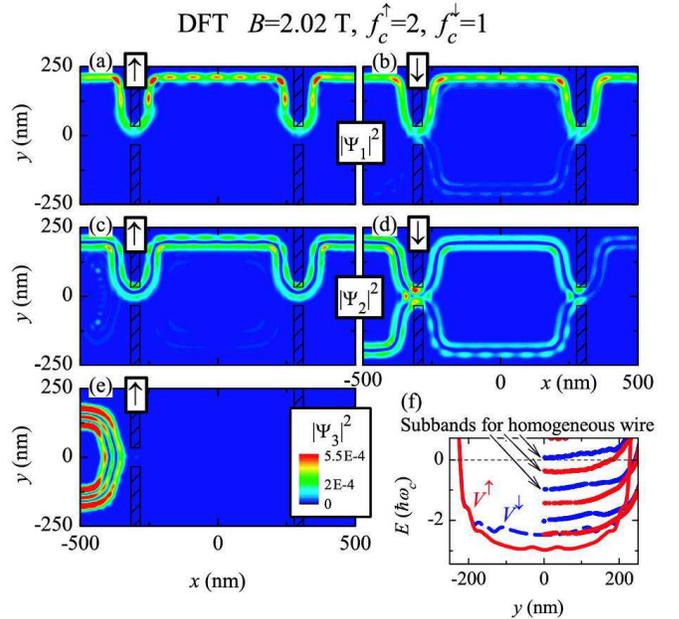} 
\caption{(color online) The wave functions modulus at $B=2.025$ T calculated
within the spin-DFT approximation $f_{c}=f_{c}^{\uparrow}+f_{c}^{%
\downarrow}=3\ (f_{c}^{\uparrow}=2, f_{c}^{\downarrow}=1)$. Left and right
columns correspond to the spin-up and spin-down electrons. For a given
magnetic fields there are five propagating states in the leads (as
illustrated in the band diagram showed in (f)), three spin-up states and two
spin-down states. Panels (a)-(e) show the wave functions corresponding to
these states.}
\label{fig:psi_DFT}
\end{figure}

For higher magnetic fields when a number of transmitted channels in the QPC
is reduced below one, $f_{c}=0$, the electron interaction becomes strongly
pronounced leading to smearing out the AB oscillations and to emergence of a
new oscillation pattern, Fig. \ref{fig:AB_oscillations} (d). For higher
magnetic field the compressible strip forms in the center of the dot. As a
result, the electrons are scattered directly in and out of this region
instead of following well defined closed paths along the dot perimeter, see
the inset to Fig. \ref{fig:AB_oscillations} (d). This leads to the
suppression of the AB oscillations and emergence of a new pattern which
periodicity $\Delta B\approx 0.22$ T is consistent with the area of a
compressible strip inside the dot ($\sim 135\times 135$ nm$^{2}$). It should
be mentioned that the non-interacting approach (where no compressible strips
are present) always shows a perfect AB periodicity.

We also calculated the spin-resolved conductance $G^{\sigma }=G^{\sigma }(B)$
within the DFT approach with the exchange-correlation effects included in
the local spin density approximation. The spin-resolved Landau levels in the
QPC constriction depopulate one by one leading to the AB oscillations with
the same periodicity as the one calculated without the exchange-correlation
term. The corresponding wave function distributions for spin resolved
electrons for $f_{c}=f_{c}^{\uparrow }+f_{c}^{\downarrow }=3\
(f_{c}^{\uparrow }=2,f_{c}^{\downarrow }=1)$ are shown in Fig. \ref%
{fig:psi_DFT}. They also show the same features as those for spinless
interacting electrons in the Hartree approximation, namely, adiabatic
character of transport for the lowest $f_{c}$ states, little intermode
scattering as well as an absence of the compressible island in the center of
the interferometer.
\begin{figure}[tbp]
\includegraphics[keepaspectratio, width=\columnwidth]{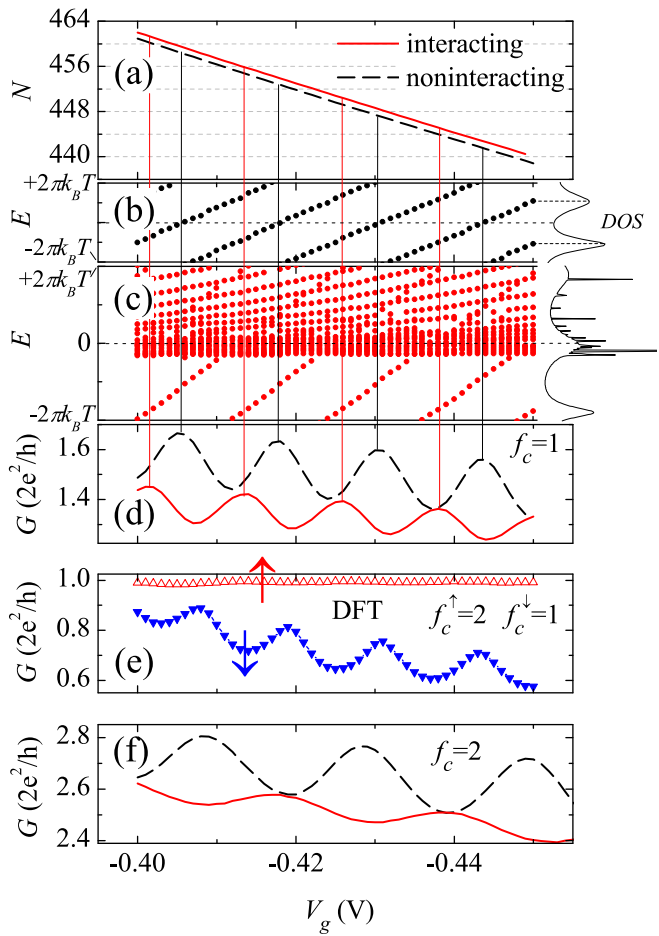} 
\caption{(color online) (a) The number of electrons inside the dot of the AB interferometer for $%
f_{c}=1$ for interacting and noninteracting spinless electrons as a function
of the gate voltage $V_{g}$. Evolution of the resonant energy levels in the
vicinity of $E_{F}$ for (b) noninteracting and (c) interacting electrons.
The insets in (b), (c) show DOS in the dot for the specified value of the
field $V_{g}=-0.45$ V; note that evolution of the energy levels was obtained
from the peak positions of the DOS at each given value of $V_{g}$.
Conductance of interacting and noninteracting spinless electrons for (d)
$f_{c}=1$ and (f) $f_{c}=2$. (e) Spin-DFT conductance of the AB
interferometer for $f_{c}=f_{c}^{\uparrow }+f_{c}^{\downarrow }=3\
(f_{c}^{\uparrow }=2,f_{c}^{\downarrow }=1)$. The magnetic field is (a)-(d) $%
B=2$ T; (e) $B=1.96$ T; (f) $B=1.25$ T.}
\label{fig:GVg}
\end{figure}

The conductance of the AB interferometer as a function of voltage applied to
the gate defining the dot, $V_{g}$, is shown in Fig. \ref{fig:GVg} for
spinless electrons for $f_{c}=1,2$, as well as for the spin-resolved case $%
f_{c}=3$ $(f_{c}^{\uparrow }=2,f_{c}^{\downarrow }=1).$ The DOS of
interacting electrons, as in the case when magnetic field was varied, shows
pinning of the weakly coupled narrow resonant states (situated inside the
dot) to the Fermi level. Also, as in the case when $B$ was varied, the
conductance of noninteracting and interacting electrons (both
spin-degenerate and spin-resolved) is practically the same. Every time a
broad resonant level crosses $E_{F},$ the conductance exhibits maximum (see
Fig. \ref{fig:GVg} (b), (c)). The ratio of the periods of the oscillations
for both $f_{c}=1$ and $f_{c}=2$ is fully consistent with the conventional
AB formula (\ref{period_Vg}), $\Delta V_{g}^{1}/\Delta V_{g}^{2}=B^{2}/B^{1}$
(note that the dot area varies approximately linearly with variation of the
gate voltage of the side gate, $\Delta S=\alpha \Delta V_{g}$\cite%
{open_dot_PRB}; indexes 1,2 correspond to $f_{c}=1,2$). Finally we notice
that even though each AB resonance is mediated by a single level, the number
of electrons in the dot between two consecutive peaks decreases by more than
one, see Fig. \ref{fig:GVg} (a). These electrons are those that depopulate
the states inside the dot (narrow resonances in the DOS, Fig. \ref{period_Vg}
(c)) and thus are not manifest in the conductance.

\section{Discussion.}

\label{sec:discussion}

The results of the conductance calculations based on the Hartree and
spin-DFT approaches presented in the previous section for $f_{c}\geq 1$ are
in excellent agreement with the conventional AB formula, Eq. (\ref{period_1}%
) predicting the same periodicity regardless of the number of the fully
transmitted channels $f_{c}$. This is in obvious disagreement with the
experiments\cite{Goldman_2005,Goldman_2007} that show a deviation from the
conventional AB periodicity by a factor of $1/f_{c},$ Eq. (\ref{period_f_c}%
). Besides, the AB calculated periodicity as a function of the gate voltage $%
V_g$ is consistent with the conventional AB formula (\ref{period_Vg}), which
also contradicts to the experimental findings showing the same periodicity
for all $f_c$ independent of the magnetic field. In our discussion of the
validity of the present approach, we argued that, as far as the conductance
is concerned, the present method is justified for the case of the strong
coupling when the conductance of the device exceeds the conductance unit, $%
G>G_{0},$ such that the electron number inside the structure is not expected
to be quantized. However, our calculations for $f_{c}\geq 1$ do not recover
the experimental conductance even thought the total conductance of the
system exceeds $G_{0}.$ What is the reason for this discrepancy?

We argue that inability of the \textquotedblleft standard
approach\textquotedblright\ to recover the experimental periodicity is an
indirect evidence that, even though $G>G_{0}$, the electron charge in the
interferometer is quantized and thus the Coulomb blockade effects become
dominant in the conductance. This is because of the adiabatic character of
the transport when the lowest $f_{c}$ states pass thought the interferometer
with the transmission coefficient close to one, see Figs. \ref{fig:psi},\ref%
{fig:Tij},\ref{fig:psi_DFT}. The highest state passing through the QPCs, $%
f_{c}+1$, (giving rise to the AB oscillations in the transition regions
between the plateaus) becomes thus effectively decoupled from the remaining $%
f_{c}$ states that pass through the interferometer practically without
reflection. Therefore the AB interferometer effectively confines only
electrons belonging to the highest ($f_{c}+1$) subband passing through QPC.
Because the conductance of this state is always smaller than one, the dot is
in the weak coupling regime, even though the total dot conductance is larger
than $G_{0}$ (due to the lower $f_{c}$ states that pass adiabatically
through the interferometer). As a results, the electron charge inside the
dot becomes quantized and transport through the interferometer becomes
strongly affected by the Coulomb blockade effect.

Note that manifestation of the Coulomb blockade effects in the conductance
of open dots is not limited to the edge state regime only. Liang \textit{et
al.}\cite{Liang} demonstrated that the adiabatic transport regime can be
achieved in an open quantum dot even at zero field leading to the dot
conductance being dominated by combined charging and ballistic transport
within a wide range $0<G<6e^{2}/h.$ The dot of Ref. \onlinecite{Liang} was
designed such that the intermode scattering was practically absent. As a
result, the lowest $f_{c}$ states propagated through the dot adiabatically
with very little reflection, whereas highest states with the transmission $%
T<1$ gave rise to the Coulomb blockade effects in the conductance.

We argued in Sec. \ref{sec:Model} that while the present \textquotedblleft
standard approach\textquotedblright\ is not expected to describe
single-electron tunneling effects (leading to the Coulomb blockade peaks in
the conductance), one can expect that it correctly reproduces a global
electrostatics of the interferometer and microscopic structure of the
quantum mechanical edge states regardless whether the conductance is
dominated by a single-electron charging or not. In its turn, such
information can be a basis for phenomenological models aiming at description
of the effects of single-electron charging in the AB interferometer.

Such model calculations have been recently reported by Rosenow and Halperin
who studied the effect of the single electron charging on the periodicity of
the AB interferometer\cite{Rosenow}. In the absence of detailed microscopic
picture of the edge state structure in the interferometer, the authors
considered several possible scenarios of coupling between the edge states in
the leads and states in the island. The important feature of their model was
the presence of the compressible region in the center of the island. Our
macroscopic calculations however do not support the assumption of the
formation of the compressible strip inside the interferometer. Our
calculations for $f_{c}\geq 1$ demonstrate only the onset of the formation
of the compressible region where just a few resonant levels of the dot
become pinned to the Fermi energy. The formation of the compressible region
inside the interferometer occurs at larger fields corresponding to $f_{c}=0$
(where however the conductance is dominated by the single-electron effects
anyway because $G<G_{0}).$ Note that Dharma-wardana \textit{et al.}'s model%
\cite{Dharma-wardana} of single electron charging in the open AB
interferometer does not seem to rely on the presence of the compressible
island inside the dot.

A microscopic picture emerging from our calculation can be summarized as
follows.

(\textit{i}) The lowest $f_{c}$ states pass through the QPC almost
adiabatically contributing very little to the conductance oscillations (see
Fig. \ref{fig:psi} (d) and Fig. \ref{fig:psi_DFT} (a)-(c)).

(\textit{ii}) The state $f_{c}+1$ passes through the interferometer with the
transmission probability $0<T<1$ giving rise to the transition region
between the conductance plateaus that is modulated by the AB oscillations.
Inside the interferometer this state retains its edge-state character (see
Fig. \ref{fig:psi} (e) and Fig. \ref{fig:psi_DFT} (d)). The AB oscillations
are related to excitation of the resonant states of the dot that are
situated close to the dot boundary and thus are strongly coupled to the
leads. These states are manifest in the density of states as relatively
broad peaks with broadening $\Gamma \sim kT$ (see Figs. \ref{fig:resonances},%
\ref{fig:LDOS},\ref{fig:GVg}).

(\textit{iii}) The $f_{c}+1$ state (and, to a lesser extend, all lowest $%
f_{c}$ states) also excite very narrow resonant states with broadening $%
\Gamma \ll kT$ situated inside the island and thus weakly coupled to the leads (see Figs.
\ref{fig:resonances},\ref{fig:LDOS},\ref{fig:GVg}). These states are pinned to the Fermi
energy and the excitation of these states corresponds to the onset of formation of the
compressible island inside the interferometer. Note however that compressible island
inside the
interferometer forms at much larger fields, see inset to Fig. \ref%
{fig:AB_oscillations} (d). Because both broad and narrow states correspond
to the addition (or subtraction) of one electron to (or from) the dot, both
of them can contribute to single-electron charging giving rise to
modification of the conventional AB periodicity according to Eq. (\ref%
{period_f_c}).

(\textit{iv}) Finally, the $f_{c}+2$ state is almost completely reflected by
the QPC, see (see Fig. \ref{fig:psi} (f) and Fig. \ref{fig:psi_DFT} (e)).
This state might or might not form a compressible strip in the leads
(depending on whether it is respectively the last filled LL or not).
However, because of the weak coupling to the states in the dot, the
compressibility of this state has a little significance for the transport
through the interferometer.

Finally we stress that our approach corresponds to the coherent electron
transport through the interferometer. It does not account for incoherent
processes such as spin flips and interlevel scattering that might lead to
redistribution of electrons between outer (broad) and inner (narrow)
resonant states in the dot. Such electron transfer between different LLs is
shown to influence an addition spectrum of a closed (strongly
Coulomb-blockaded) dot\cite{McEuen}. However, in the case of open dot
considered in this study it is not clear whether such processes would
significantly affect the conductance of the interferometer, because the
dwell time of the electrons in the open dot might be much smaller that
inelastic scattering time associated with the interlevel relaxation.

\section{Conclusion.}

We provide a microscopic picture of the quantum transport in the
Aharonov-Bohm interferometer taking into account electron interaction within
the Hartree and the spin-DFT approximations. We discuss the structure of the
edge states for different numbers of the Landau levels in the leads,
structure of the states in the dot, coupling between the states in the dot
and the leads, and the formation of compressible/incompressible strips in
the interferometer. We discuss the applicability of our approach and argue
that it provides a reliable description of a global electrostatics of the
interferometer and a microscopic structure of the quantum mechanical edge
states and coupling between them. On the other hand, the present approach is
not expected to reproduce the conductance in the weak coupling regime of the
Coulomb blockade, if the electron number inside the interferometer becomes
quantized. We compare our conductance calculation with the experiment\cite%
{Goldman_2005,Goldman_2007} and argue that the inability of the present
approach to reproduce the unexpected periodicity of the AB oscillations, Eq.
(\ref{period_f_c}), can be taken as an indirect evidence that this
periodicity is caused by the Coulomb blockade-type effects.

Our transport calculations thus demonstrate that an accurate description of the
conductance of the AB interferometer would require theories that go beyond the
\textquotedblleft standard approach\textquotedblright \cite{Koentopp} based on the
ground-state DFT in the Landauer formula that was utilized
in the present paper. Such the theories (as e.g. reported in Refs. %
\onlinecite{Dharma-wardana,Rosenow}) essentially rely on specific phenomenological models
of the states in the leads and in the central island and their coupling in the
interferometer. Our work, therefore, provides a basis for such the theories, giving a
detailed microscopic description of the propagating states and the global electrostatics
in the system at hand. Such a microscopic description is summarized in Sec.
\ref{sec:discussion}. In particular, our findings does not directly support the model of
Rosenow and Halperin that relies on the existence of the compressible island inside the
interferometer and its coupling to the leads. Our findings thus indicate that an accurate
explanation of the unexpected periodicity of the AB oscillations might need exploring
alternative theories based on the microscopic picture of interesting electrons developed
in the present paper.

\textit{Acknowledgement}. We are thankful V. Goldman for drawing our
attention to the current problem. We appreciate valuable discussions with V.
Goldman, T. Heinzel, A. Sachrajda. We acknowledge access to computational
facilities of the National Supercomputer Center (Link\"{o}ping) provided
through SNIC.

\end{document}